\documentclass[%
 aip,
 amsmath,amssymb,floatfix,
 reprint,%
]{revtex4-1}

\usepackage{graphicx}% Include figure files
\usepackage{dcolumn}% Align table columns on decimal point
\usepackage{bm}% bold math
\usepackage[utf8]{inputenc}
\usepackage[T1]{fontenc}
\usepackage{mathptmx}
\usepackage{etoolbox}

\makeatletter
\def\@email#1#2{%
 \endgroup
 \patchcmd{\titleblock@produce}
  {\frontmatter@RRAPformat}
  {\frontmatter@RRAPformat{\produce@RRAP{*#1\href{mailto:#2}{#2}}}\frontmatter@RRAPformat}
  {}{}
}%
\makeatother

\begin{document}

\preprint{APS/123-QED}

\title{Hybrid Magnonic Reservoir Computing}% Force line breaks with \\

\author{Cliff B. Abbott}
\author{Dmytro A. Bozhko}%
\affiliation{%
 Department of Physics and Energy Science, University of Colorado Colorado Springs, Colorado Springs, 80918 CO, USA
}%

\date{\today}% It is always \today, today,
             %  but any date may be explicitly specified

\begin{abstract}
Abstract: Magnonic systems have been a major area of research interest due to their potential benefits in speed and lower power consumption compared to traditional computing.  One particular area that they may be of advantage is as Physical Reservoir Computers in machine learning models.  In this work, we build on an established design for using an Auto-Oscillation Ring as a reservoir computer by introducing a simple neural network midstream and introduce an additional design using a spin wave guide with a scattering regime for processing data with different types of inputs.  We simulate these designs on the new micro magnetic simulation software, Magnum.np, and show that the designs are capable of performing on various real world data sets comparably or better than traditional dense neural networks.
\end{abstract}

%\keywords{Suggested keywords}%Use showkeys class option if keyword
                              %display desired
\maketitle

%\tableofcontents

\section{\label{sec:level1}Introduction}

Reservoir computing (RC) is a machine learning concept where a lower dimensional input is passed to a system which maps to a higher dimensional output via a complex, non-linear, but dynamically consistent process.  Unlike other machine learning models, the training process is regulated to the output, or readout, layer(s) of the RC allowing for significant reductions in training time \cite{RC_Intro}.  However, the implementation of a RC can still be quite complicated and computationally expensive when done with simulated neurons, like in the Recurrent Neural Network approach.  Because of this, Physical Reservoir Computers (PRC), where the complex, nonlinear dynamics are done by exploiting natural systems dynamics rather than simulated, have become a popular area of research \cite{nakajima2020physical}.  Of the many physical systems available, magnonics is a promising candidate for next generation computers due to the potentially fast computational speed, low energy consumption compared to other electronics, and small physical size of the systems \cite{Intro1,Intro2,Intro3,Intro4,Intro5,Intro6,Intro7,Intro8,Intro9}.  While much of the current work has been focused on the implementation of spinwave based logic gates \cite{Logic1,Logic2,Logic3,Logic4,Logic5,Logic6,Logic7,Logic8}, there have recently been several proposed designs for magnonic RCs \cite{RC_Example_1,RC_Example_2,NewRC,NewRC2}. 

In this work, we build on the magnonic auto-oscillation ring (AOR) proposed in \cite{watt2021implementing} and model through numerical simulation \cite{watt2024numerical} by Watt \textit{et al}. by adding a dense neural network midstream with the new micromagnetic simulation package Magnum.np \cite{Magnum}.  Also, we explore differences between encoding input information into either the amplitude or phase of excited spin waves.  Previous work has primarily been focused on proving the design against theoretical benchmarks such as input memory and parity memory.  Here we use historical S\&P stock market data to test the design on real world applications.

In addition to the AOR, we introduce a design for a spin-wave-based system for feature mixing we call a Parallel Input Scattering Model (PSM).  The PSM encodes features into two separate spin waves that are simultaneously sent through a scattering region before being read out.  We tested various capabilities of this model on the Iris and Statlog datasets from the UCI Machine Learning Repository \cite{IRIS,statlog} and on a 3rd custom dataset.  

The rest of the paper is structured as follows.  Section II: Methods covers (A) the setup of the simulation software, followed by (B) the design and operation of the AOR as well as the data used and it's preprocessing, and the final subsection (C) covers the same material for the PSM.  Section III: Results reports the various testing results for (A) the AOR and (B) the PSM.  Finally, Section IV: Conclusions covers the insights from the simulations and suggestions for future work.

\section{Methods}
\subsection{Simulation}
For simulating the designs in this paper, we used the Magnum.np Python package.  This package is built on the PyTorch backend allowing python native simulations.  It also allows for seamless integration of Torch machine learning models, back propagation methods, and built in GPU optimization methods.  For all simulations, a grid size $2.5\,\mathrm{nm^{3}}$, saturation magnetization value of $140\,\mathrm{kA/m}$, exchange constant of $3.5\,\mathrm{pJ/m}$, and Gilbert damping constant of $0.0002$ were used to approximate a Yttrium Iron Garnet thin film.  Performance criteria for all simulations is the percentage accuracy at identifying the correct answer.  The simulations were done on an Nvidia DGX Workstation utilizing four A100 8GB GPUs.
\subsection{Auto Oscillation Ring}
\begin{figure}
\includegraphics[width=0.4\textwidth]{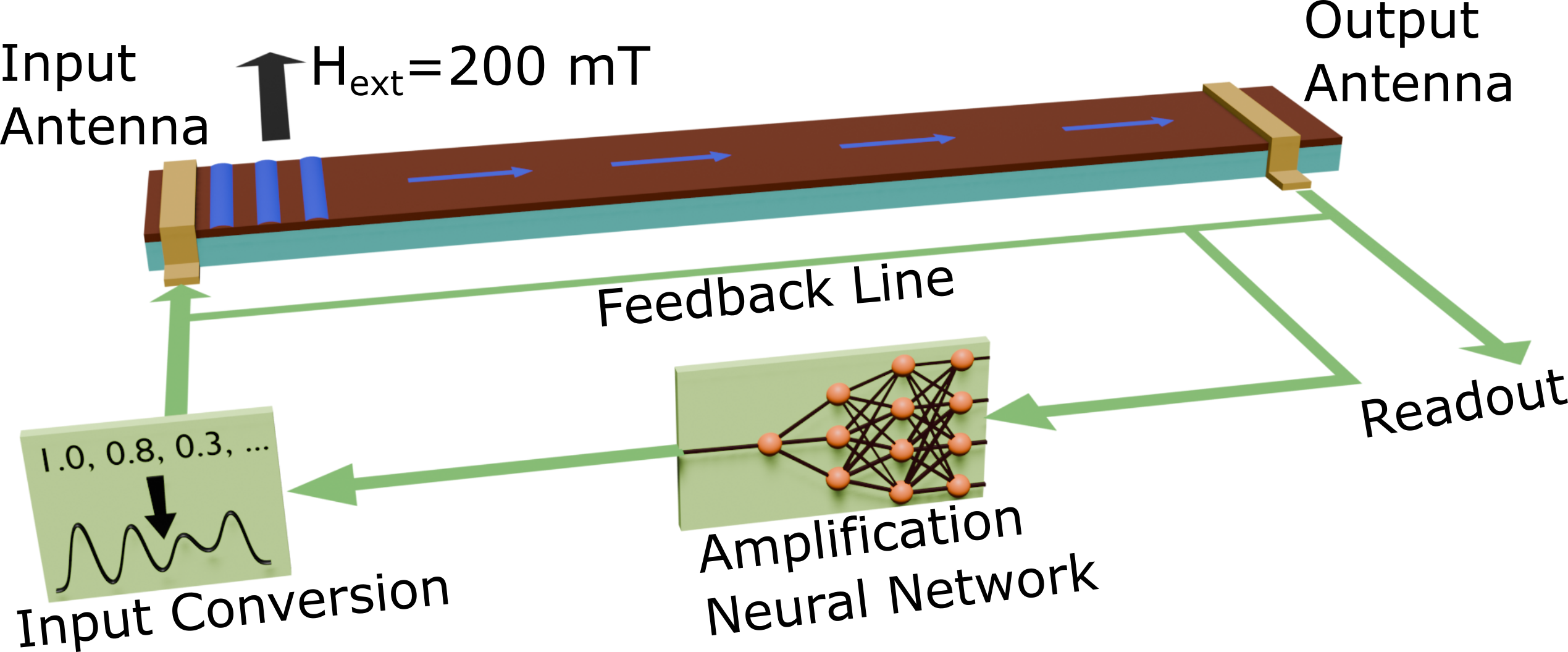}
\caption{\label{fig:AOR} Layout of the AOR design used.  The data is encoded into the input signal which is then used to excite magnetostatic surface waves which propagate down the wave guide.  The output antenna converts the spin wave back into an electronic signal which is then sent to 3 different channels: the readout, the feedback line, and the amplification neural network.  The feedback line adds the output signal to the incoming signal at the input antenna.  The amplification neural network takes the output from the previous interval and uses it to determine some additional amplification to the next intervals input signal. }
\end{figure}

\begin{figure}
\includegraphics[width=0.4\textwidth]{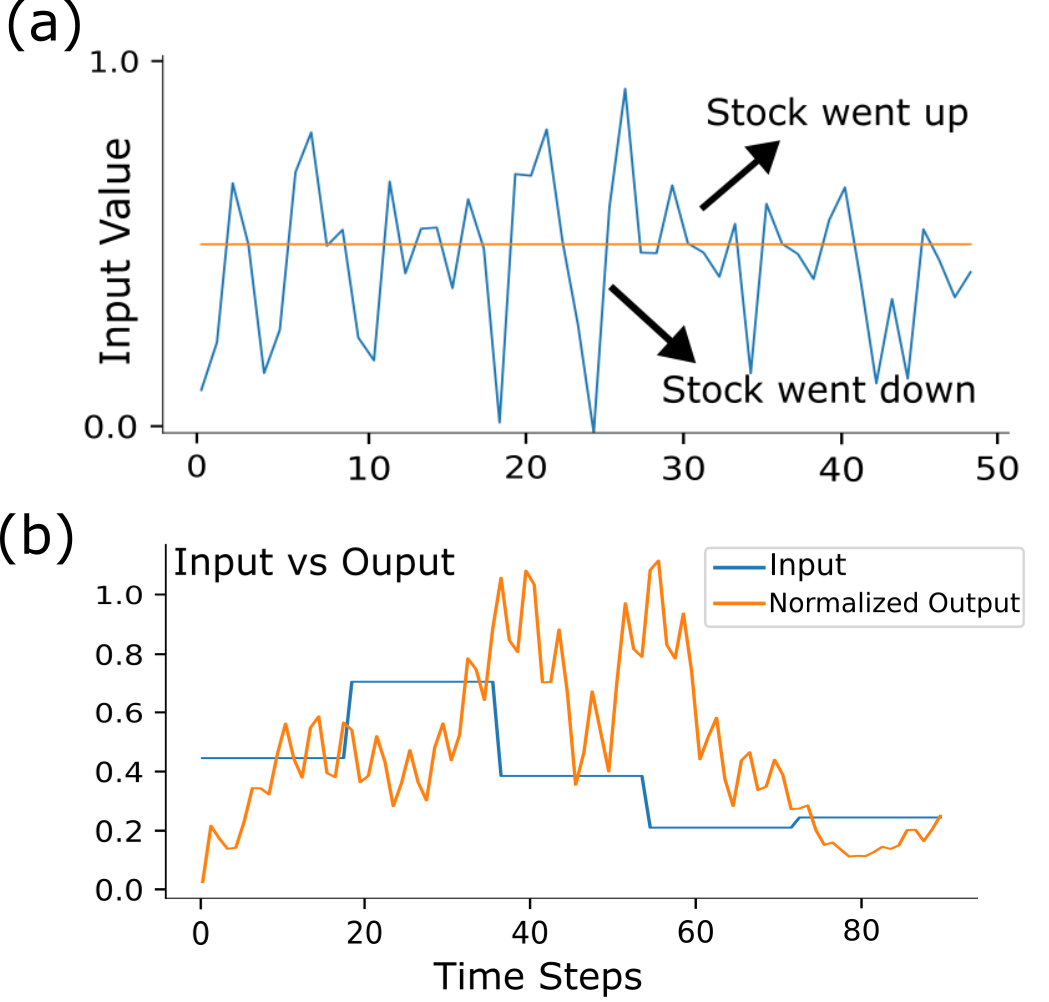}
\caption{\label{fig:stocks} a) The result of the preprocessing on a sample of the stock data. b) A comparison of the output behavior to the input values.  Output values  normalized to be comparable to the input.}
\end{figure}

Figure \ref{fig:AOR}
 shows the layout of the AOR.  A grid size of $1250\times50\times2.5\,\mathrm{nm}$ was used for this simulation.  A $200\,\mathrm{mT}$ external bias field was initialized in plane and perpendicular to the direction of propagation for the waveguide and magnetostatic surface waves were excited at the input antenna at a frequency of $14\,\mathrm{GHz}$ for $0.3$ nanoseconds corresponding to 4 periods per input signal.  The output is read per $1/100th$ of a nanosecond into the amplification neural network (ANN) such that the output $a$ of the network with inputs $x$ and at input interval $i$ is $a_i=(x_{i-1,0},x_{i-1,1}, ... ,x_{i-1,30})$.  The ANN is a 2 layer multi-layer perceptron (MLP) with a first layer of 10 nodes and ReLU activation, followed by a single node layer and sigmoid activation.  The sigmoid activation was used to ensure that the output values were between 0 and 1 so as to prevent the ANN from blowing up the input signal amplitude to unrealistic values.  For analysis, the input signal before amplification by the ANN, the signal from the output antenna, and the difference between the two are used.
\\
Watt \textit{et al}. \cite{watt2021implementing,watt2024numerical} used a test measure of input memory (recover the input value j inputs ago) and a parity memory (the binary sum of the previous j inputs).  Here we used the closing values for 250 consecutive trading days of the US S\&P Stock Index taken from Yahoo! Finance.  The data was pre-processed into a percentage move from the previous day's closing value and re-scaled to a maximum and minimum of 1 and 0 respectively (Fig \ref{fig:stocks}a).
  The target value was 1 if the next day was a positive move or 0 if it was negative.  The inputs $I$ were then encoded into input signal $S$ as either a modulation of the amplitude $S=I\times Sin(2\pi ft)$ or of the phase $S=Sin(2\pi ft + \pi I)$.  The output of the simulation was additionally processed in one of three ways:
\\
$\bm{1.}$ A direct linear mapping
\begin{center}
    $Weights = Train Output^{-1}\cdot TrainTargets$\\
    $Prediction = Weights \cdot Test Output$\\
    \end{center}
    % \begin{FlushLeft}
    $\bm{2.}$ Using an ensemble of linear maps approach   
    % \end{FlushLeft}
  \begin{center}
  $TrainOutput \rightarrow (Train Out_{1}, ... , Train Out_{N})$\\
    $Weights_{1} = TrainOut_{1}^{-1}\cdot TrainTargets_{1}$\\
    ...\\
   $Weights_{N} = TrainOut N^{-1}\cdot TrainTargets_{N}$\\
    $Prediction = \frac{1}{N} \sum_{N}{Weights_{N} \cdot TestOutput}$
  \end{center}
% \begin{FlushLeft}
    $\bm{3.}$ An additional small MLP.
% \end{FlushLeft}

\subsection{Parallel Input Scattering Model}
\begin{figure}
\includegraphics[width=0.4\textwidth]{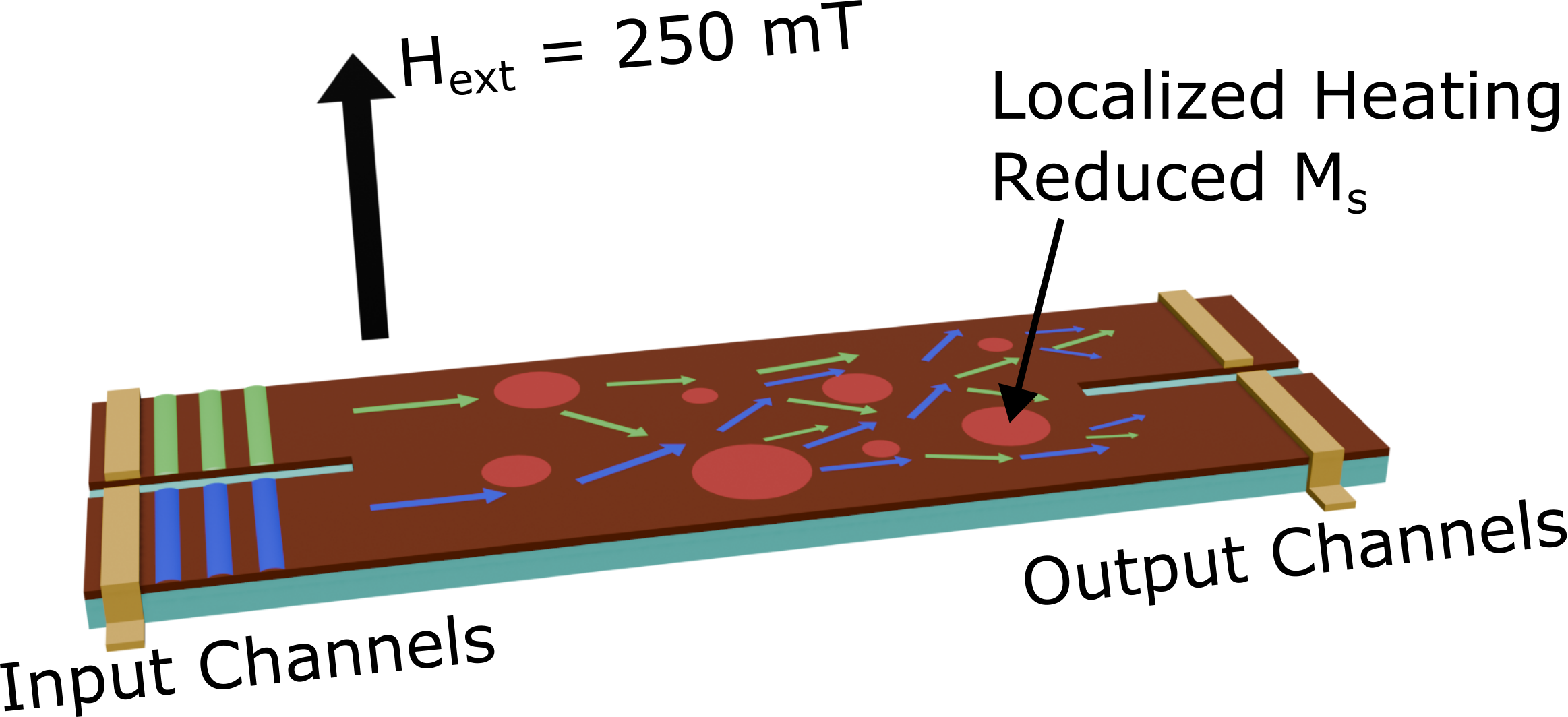}
\caption{\label{fig:2RP} Layout of the PSM design used.  The input region is divided into 2 input channels.  Spin waves are excited in the input channels propagating down to the scattering region where the two waves mix.  The output region was divided into 1, 2, or 3 read out channels (2 readout channels shown here).}
\end{figure}

The PSM (Fig \ref{fig:2RP}) was a $1250\times50\times2.5\,\mathrm{nm}$ grid with a slice at the excitation side cutting two channels into the wave guide.  At the output, the region was divided into 1, 2, or 3 output channels.  
A $250\,\mathrm{mT}$ external bias field was applied out of plane and the input antennas excited forward volume magnetostatic spin waves at 6.2 GHz.  Forward volume wave geometry was chosen for this design so that waves could propagate isotropically in the scattering region.  This allows the input waves from the different channels to scatter into each other.  The interaction and interference of the scattered waves is analogous to the non-linear connectivity in a dense neural network layer.  The scattering was achieved by applying a reduced saturation magnetization in randomized spots as would be seen through local laser heating \cite{magGrad}.  Multiple data sets were used to test the PSM in different capacities.  In all cases, the neural network to analyze the outputs was a small 1 hidden layer MLP with varying node counts.

\begin{figure}
\includegraphics[width=0.4\textwidth]{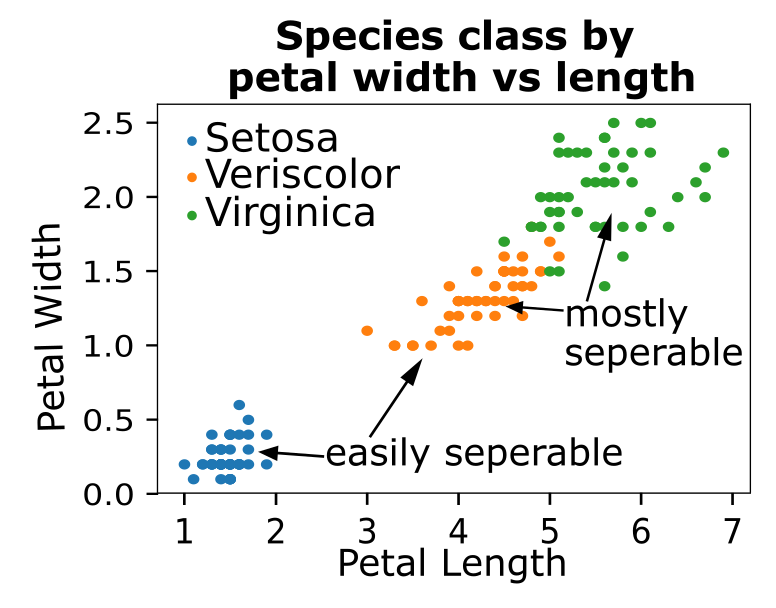}[a]
\includegraphics[width=0.4\textwidth]{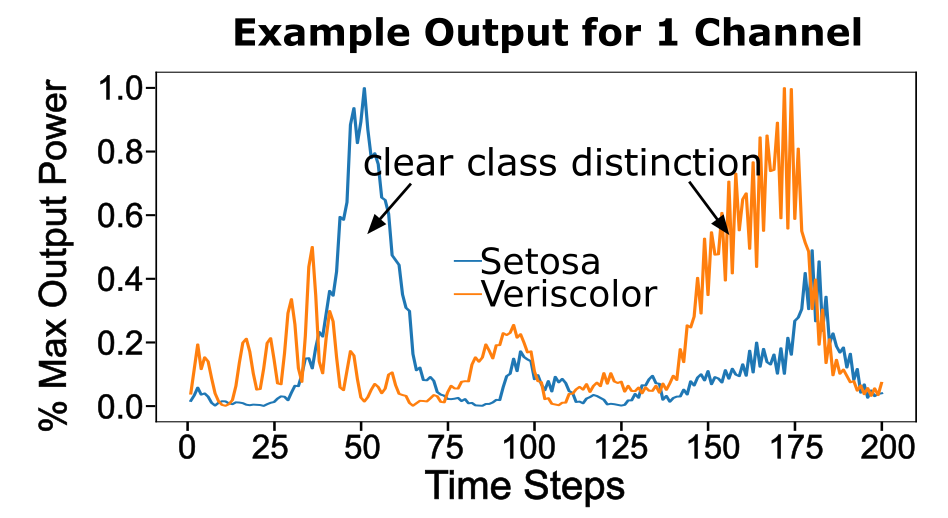}[b]
\caption{\label{fig:id} a) Iris classes plotted by petal length and width.  The Setosa and Veriscolor classes make up the easily separable set.  The Veriscolor and Virginica classes make up the mostly separable set. b) The scaled readout from the 1 output channel simulation for two samples within the easily separable set.}
\end{figure}

The initial data used was the Iris dataset from the UCI Machine Learning Repository \cite{IRIS}.  The data has 3 classes of species (Setosa, Veriscolor, and Virginica) and 4 features for each class (petal length, petal width, sepal length, and sepal width).  Using only 2 features, the classes can be broken up into two sets of 2 classes, one easily separable and one only mostly separable (Fig \ref{fig:id}a).  As with the stock data and all additional data sets, the features were scaled to a maximum value of 1.  The width values were then inverted by subtracting their value from 1.  This was done to prevent the PSM from simply being able to tell the classes apart simply by the output amplitude as both the width and length of the petals and sepals increased in value from $Setosa \rightarrow Veriscolor \rightarrow Virginica$.  Inverting one of the feature measures forces the PSM to differentiate the classes through the scattering process.  The intention of this test was to gauge the potential of the PSM for feature mixing before continuing to more complicated structures.

The second data set was the Statlog (German Credit Data) set also from the UCI Machine Learning Repository \cite{statlog}.  This data set contains 24 features, 12 of which are binary and 12 of which are numerical.  Only the 12 numerical features were used as inputs for the PSM.  The labels were set to 0 for good credit risk and 1 for bad credit risk. The inputs were stacked into a 6x2 tensor and sent into the 2 PSM input channels 1 at a time for 6 consecutive intervals.  Only the 1 output channel PSM was used for this test.  The intention of this test was to push the feature mixing and information preservation capability on a single PSM.  

The final data set, referred to here as the Dimensional Reduction set, was a created dataset with 4 features and 3 classes based on a randomly chosen function of those features.  The features were broken up into sets of 2 and each set sent to their own PSM with a single output channel.  The output of the two initial PSMs were fed into the input channels of a final PSM, also with a single output channel.  The intention of this test was to gauge the capability of the PSM for dimensional reduction of data channels.

\section{Results}
\subsection{Auto Oscillation Ring}
To get a representation of the model's abilities, each output method (linear map, ensemble, or neural network) was performed 1000 times each on a randomized selection of the data determined using the SKlearn Train Test Split module with splits of 10, 20, 30, 40, and 50 percent.  This resulted in 5000 trials for each method and a total of 15000 trials.  The direct linear mapping and ensemble approaches were nearly identical and never better on the AOR output than on the data directly.  Improvements were seen using the neural network method on the AOR output compared to a similar network fed the data directly.  As shown in Table \ref{tab:AOR Results}, 
\begin{table}
\begin{ruledtabular}
\textbf{Auto Oscillation Ring}\\
\textbf{Neural Network Method Results}\\
\begin{tabular}{cccc}
 &Max&Mean&Std\\
\hline
Reference - No AOR (NN)&80.0&61.6&11.75\\
Reference - No AOR (LM)&80.0&62.7&5.41\\
\hline
Amplitude Encoded - (NN)&88.0&63.2*&8.19\\
Amplitude Encoded - (LM)&84.0&56.7&6.23\\
\hline
Amplitude Encoded - ANN (NN)&92.0&64.7*&6.58\\
Amplitude Encoded - ANN (LM)&84.0&57.5&6.16\\
\hline
Phase Encoded - (NN)&88.0&61.7&9.10\\
Phase Encoded - (LM)&84.0&54.5&6.52\\
\hline
Phase Encoded - ANN (NN)&88.0&61.8&9.12\\
Phase Encoded - ANN (LM)&88.0&55.3&6.52\\
\end{tabular}
\end{ruledtabular}
\caption{\label{tab:AOR Results}%
AOR Results: NN - Neural Network output analysis.  LM - Linear Mapping output analysis.  ANN - Amplification Neural Network was included.  The results show that the inclusion of the ANN generally improved results and that amplification encoding produced the best results.  *indicates an improvement over reference numbers.}
\end{table}phase encoding had a minimal impact on performance whether or not the ANN was included.  There was, however, a statistically significant improvement in performance with amplitude encoding.  Additional improvement was also seen in all cases when the ANN was included.  The AOR with ANN and amplitude encoding had the highest maximum accuracy, highest mean accuracy, and lowest standard deviation.

We also found that training of the ANN was not possible.  When using back propagation methods, the parameter gradients were always zero.  To verify this result and ensure that the zero gradient was not due to a mistake in the code detaching the gradient during simulation, we also attempted training using the parameter shift method popular in quantum machine learning \cite{ParamShift}.  This method simply uses the derivation of a derivative $(f(x+\Delta)-f(x))/\Delta$ which in the limit $\Delta\rightarrow0$ is $\frac{df}{dx}$.  It requires 1 additional forward run of the model per parameter but is not susceptible to the detachment error.  The zero gradient of the ANN does fit within the RC theory as the method of non-linearity introduced into the system is not relevant, only the amount and consistency.
\subsection{Parallel Input Scattering Model}
\subsubsection{Iris Data}
\begin{figure}
\includegraphics[width=0.4\textwidth]{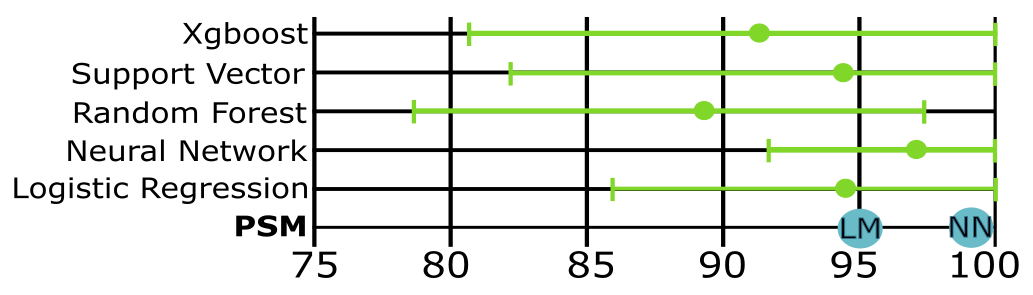}
\caption{\label{fig:PSM_iris_comp} Benchmark results taken from \cite{IRIS} with PSM results added for comparison.  LM - Linear Mapping result (95.0\%).  NN - Neural Network result (98.3\%).}
\end{figure}

The Iris data represents an entry level test to determine if the PSM was worth pursuing.  The design performed quite well showing promise for further testing.  The PSM was able to clearly distinguish classes in most cases (see for example Fig \ref{fig:id}b).  Inverting the data along one axis (flipping Fig \ref{fig:id}a horizontally) significantly improved results suggesting that the amplitude ratio between the two input channels was more important than the overall amplitude.  Results for the inverted axis case are presented in table \ref{tab:PSM_iris_results} and a comparison to other machine learning methods (fig \ref{fig:PSM_iris_comp}) shows that the PSM performed extremely well even for such a basic task.

\begin{table}
\begin{ruledtabular}
\textbf{Parallel Input Scattering Model Results}\\
\textbf{Iris Data Accuracy}\\
\begin{tabular}{cccc}
Output Channels&1&2&3\\
\textbf{Setosa/Veriscolor}(a)\\
Linear Map&97.5&100&100\\
Neural Network&100&100&100\\
\textbf{Viginica/Veriscolor}(b)\\
Linear Map&85.0&90.0&90.0\\
Neural Network&95.0&97.5&97.5\\
\textbf{Setosa/Veriscolor/Viginica}(c)\\
Linear Map&90.0&95.0&95.0\\
Neural Network&92.0&98.3&97.1\\
\end{tabular}
\end{ruledtabular}
\caption{\label{tab:PSM_iris_results}%
Accuracies for the Iris data based on the number of output channels. (a) the easily distinguishable classes. (b) the mostly distinguishable classes. (c) distinguishing between all 3 classes at the same time.  Standard deviation for all results was less than 1\%.}
\end{table}

\subsubsection{Statlog Data}
\begin{figure}
\includegraphics[width=0.4\textwidth]{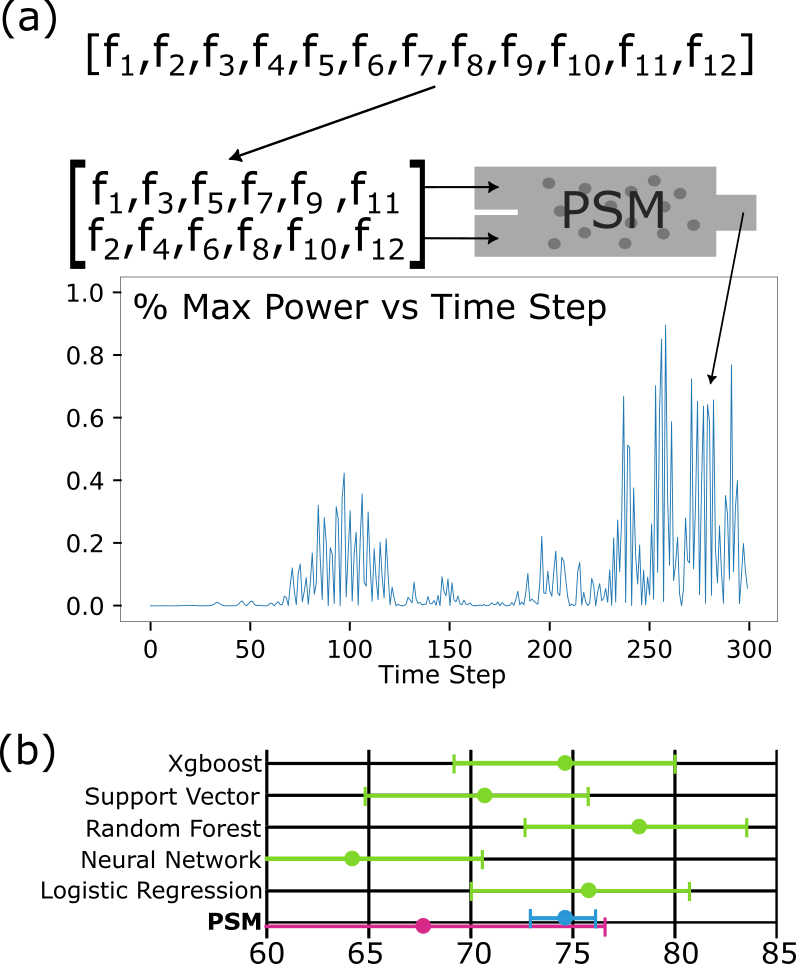}
\caption{\label{fig:PSM_statlog}(a) Features are broken up into pairs. Each pair is sent to the PSM one at a time. 
 (b) Benchmark results taken from \cite{statlog} with PSM results added for comparison.  Red Line - Linear Mapping result (67.7 +/- 9.6\%).  Blue Line - Neural Network result (74.7 +/- 2.0\%).}
\end{figure}
The Statlog data demonstrated that the PSM was able to preserve information from multiple features put into the system sequentially (Fig \ref{fig:PSM_statlog}a).  The waves that were scattered continue to bounce around and effect the output channel well after the initial wavefront has passed through.  This appears to be sufficient as a form of fading memory required by RC theory.  Despite the fact that the feature data had no temporal relation, the sequential input worked well to maintain the information at the output channel.  Because the benchmark performances from \cite{statlog} are assumed to be done on the full set of 24 features, reference benchmarks were done using the same subset of 12 features (Table \ref{tab:PSM_statlog_results}).
\begin{table}
\begin{ruledtabular}
\textbf{Parallel Input Scattering Model Results}\\
\textbf{Statlog Data Accuracy}\\
\begin{tabular}{ccc}
&Mean&Std\\
Reference Linear Mapping&71.6&1.9\\
Reference Neural Network&74.3&0.9\\
PSM Linear Mapping&67.7&9.6\\
PSM Neural Network&74.7&2.0\\
\end{tabular}
\end{ruledtabular}
\caption{\label{tab:PSM_statlog_results}%
Accuracies for the Statlog data.  Reference accuracy tests were done by performing the same output analysis directly on the 12 input features.}
\end{table}
The accuracy from direct linear mapping decreased with the PSM compared to the reference.  However, the addition of a small MLP at the output recovered the loss in accuracy and slightly improved the baseline performance.  While some information may have been lost due to inserting non-temporal data in a time series fashion, it appears that the mixing from the scattering region did more than make up for loss.  These initial results suggest that the PSM is capable of reducing multiple input features into a single output channel regardless of whether they are time series data traditionally used in RC systems or not.
\subsubsection{Dimensional Reduction Data}
\begin{figure}
\includegraphics[width=0.4\textwidth]{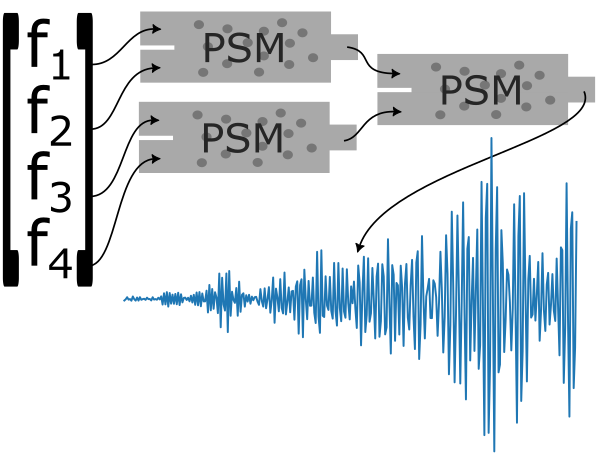}
\caption{\label{fig:PSM_DimRed}Workflow for dimensional reduction test.  Features are separated to different input channels on different devices.  Output channels from previous layer devices are passed to the input channels of the next layer until a single output channel remains.}
\end{figure}
\begin{table}
\textbf{Parallel Input Scattering Model Results}\\
\begin{ruledtabular}
\textbf{(a) Dimensional Reduction Data Accuracy}\\
\begin{tabular}{ccc}
&Mean&Std\\
Reference Linear Mapping&37.9&1.7\\
Reference Neural Network&95.7&1.3\\
PSM Linear Mapping&65.9&6.1\\
PSM Neural Network&81.2&3.8\\
\end{tabular}
\end{ruledtabular}
\begin{ruledtabular}
\textbf{(b) Accuracy On Single PSM}\\
\begin{tabular}{ccc}
&Mean&Std\\
&\textit{One Output Channel}\\
PSM Linear Mapping&66.7&5.6\\
PSM Neural Network&96.8&1.7\\
&\textit{Two Output Channels}\\
PSM Linear Mapping&69.8&5.4\\
PSM Neural Network&97.5&3.8\\
\end{tabular}
\end{ruledtabular}
\caption{\label{tab:PSM_dimred_results}%
Accuracies for the dimensional reduction data: (a) for the multiple PSM approach where only two features are provided to each initial PSM and the single channel outputs are fed into the input channels of the next layer PSM. (b) for a single PSM where all features are provided in sequential pairs.  Both one and two output channel accuracy given.}
\end{table}
The dimensional reduction data was created to test the ability of multiple PSMs to take single value features and reduce them down to single channel output channel (Fig \ref{fig:PSM_DimRed}).  Outputs from the previous layer were also amplified to be roughly the same strength as the original input layer.  The data was created by randomly choosing 3 points in a 4 dimensional space as class centers.  Additional sets of random 4 dimensional vectors were created and assigned a class based on the closest class center.  A loss of some information was seen (-14.5 $\pm$ 5.1\% accuracy) when comparing MLP accuracy's of the output channel compared to the data directly (Table \ref{tab:PSM_dimred_results} a).  There was a significant improvement in performance from direct linear mapping which is to be expected from the non-linear mixing in the RC scattering regions.  In order to confirm that the information loss was due to the design and not a peculiarity in the data, this dataset was also tested on a single PSM in the same manner as the Statlog data (results in Table \ref{tab:PSM_dimred_results} b).  The features were paired and given as inputs to the PSM sequentially.  The single PSM result was significantly better than both the multiple PSM design and the baseline reference.  This confirms that sequential passing of features on a single PSM is more effective than using multiple PSMs in a converging architecture.

\section{Conclusions}
The Auto Oscillation Ring produced a statistically relevant result for the given task.  The design showed marginal improvement with the addition of the in-line Amplification Neural Network which needed no training.  Additional improvements in the design may increase it's utility as part of a larger machine learning system.

The Parallel Input Scattering Model was designed to mimic a series of dense neural network layers by scattering a series of forward volume waves.  This design performed very well on the basic classification test with results comparable to other leading machine learning methods.  The results from the feature capacity test are arguably the most promising for future use of the PSM.  The ability of the design to convert 12 input features into 1 output channel is highly encouraging for possible use in reducing the size, complexity, and cost of large machine learning models.  The use of multiple cascading PSMs was not effective.  This is likely due to the loss in information seen from the output of the PSM without an additional neural network at the output.  This loss is carried forward from the previous PSM to the next.  

The simulation size of all designs presented was much smaller than the size of a likely physical implementation due to hardware and time constraints.  While the general physics and result trends should scale up, there may be noticeable differences when physically implementing these designs.  Future work would likely consist of building a physical system and testing for similar performance.  Should the results be consistent with the simulated results, finding a direct comparison of the speed and energy cost of magnonic systems vs traditional neural networks would be of interest.

\section{Acknowledgements}
Authors acknowledge the support from the National Science Foundation of the United States by Grant No. ECCS-2138236 .
\bibliography{Bib_export}

\end{document}